\documentstyle[epsf,aps,prl,multicol]{revtex}
\draft

\begin{document}
\title{Topological Order in the Vortex Glass Phase \\
of High-Temperature Superconductors}
\author{Jan Kierfeld$^{(1,2)}$, Thomas Nattermann$^{(1)}$, and Terence
Hwa$^{(2)}$}
\address{$^{(1)}$ Institut f\"ur Theoretische Physik der\\
Universit\"at zu K\"oln, D-50937 K\"oln, Germany \\
$^{(2)}$ Department of Physics, University of California at \\
San Diego, La Jolla, CA 92093-0319}
\date{Received: December 11, 1995}
\maketitle

\begin{abstract}
The stability of a vortex glass phase with quasi-long-range positional order
is examined for a disordered layered superconductor. The role of topological
defects is investigated using a detailed scaling argument, supplemented by
a variational
calculation. The results indicate that topological order is preserved for
a {\em wide range} of parameters in the vortex glass phase.
The extent of the stability regime is given in terms of a simple
Lindemann-like criterion.

\end{abstract}


\begin{multicols}{2}

It is well-known that the Abrikosov flux lattice in a type-II superconductor
is unstable to point disorder beyond the Larkin length \cite{L}. The nature
of the flux array at larger scales has been a subject of intense
studies~\cite{blatter}. It has been conjectured that the flux array is
{\em collectively pinned},
forming a vortex glass (VG) phase \cite{F,FGLV,natt1,FFH} with zero linear
resistivity at low temperatures. This conjecture is supported by a number
of experiments on disordered samples of high-$T_c$
superconductors~\cite{koch,gammel,safar,yeh1}, where a continuous
transition to a phase with zero
linear resistivity was found upon cooling. On the other hand,
Bitter-decoration \cite{deco}, neutron scattering \cite{neutron}, and $\mu$%
SR~\cite{msr} experiments on weakly-disordered samples have all indicated
some {\em long-range order} of the flux array, a characteristic usually
incompatible with a glass. A common interpretation for the observation of a
flux {\em lattice} is a crossover effect due to the large Larkin lengths in
weakly disordered samples. In this article, we investigate a different
possibility, that the flux array may maintain its positional long-range
order {\em much beyond} the Larkin length, in spite of pinning by
point disorders.

Such a possibility is indeed realized in a model of {\em dislocation-free}
flux line array in random media~\cite{natt1,GL}. This model is very similar
to the randomly-pinned charge-density waves and the random-field XY model
which have been studied extensively in the past decades \cite{GL,VF,CO,K}. A
variety of approximate methods have been used to obtain the conclusion that
point disorders lead to a glass phase with only logarithmic fluctuations in
the transverse displacement of the flux array. This implies the existence of
quasi-long-range positional order in the glass phase of the dislocation-free
flux array~\cite{GL}.

Recently, Giamarchi and LeDoussal suggested~\cite{GL,GL2} that such a
topologically-ordered glass may actually exist as a {\em stable}
thermodynamic phase for some range of parameters in the cuprate
superconductors. A related numerical study of the random field-XY model by
Gingras and Huse~\cite{GH} further supported this scenario. However, the
issue of spontaneous formation of topological defects (i.e., dislocation
loops) involves complicated interplay between elasticity and disorders, and has
so far not been addressed {\em quantitatively}. In this article, we
investigate this issue by studying a model of flux lines confined in the
planes of a layered superconductor [Fig.~1]. Our model allows for the formation
of dislocation loops and is amenable to analytic studies. We first
present a detailed scaling argument, which yields suppression of large
dislocation loops at finite fugacities. This result is then
supplemented by a variational calculation, from which we obtain a
Lindemann-like criterion giving the size of the stability regime
for the topologically-ordered VG. Similar behaviors are
expected in the usual experimental situation of flux lines
perpendicular to the layers.

\begin{figure}
\epsfxsize=3.2truein
\hskip 0.0truein \epsffile{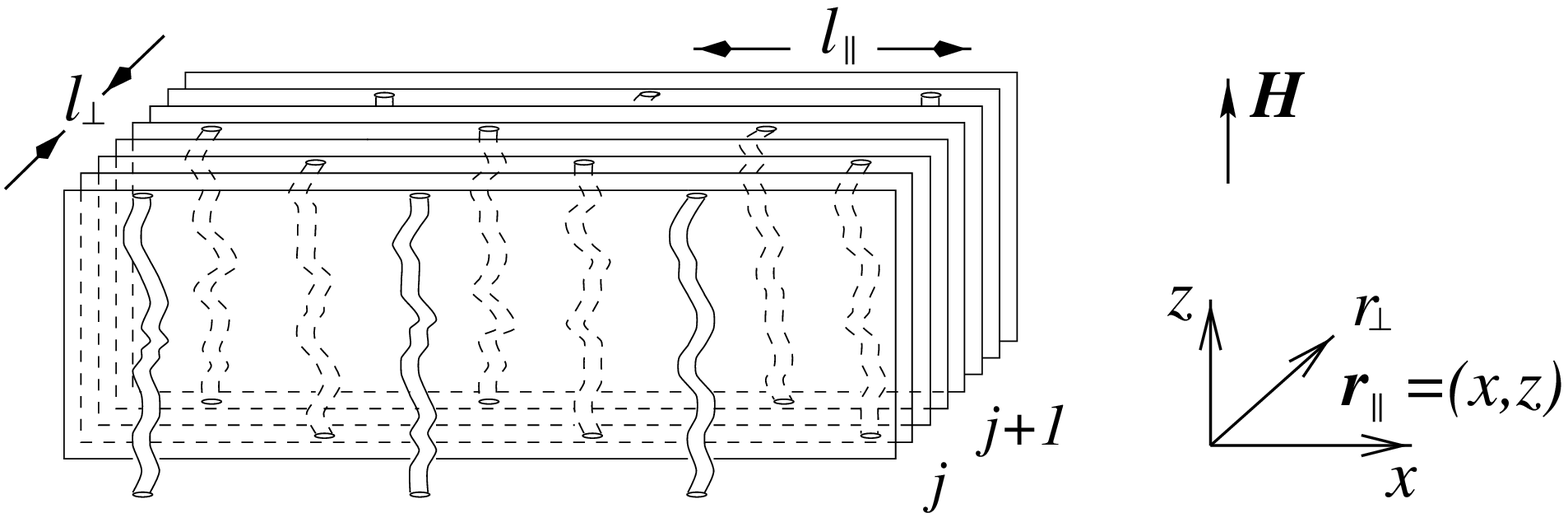}
\vspace{\baselineskip}
 { FIG 1: Flux line array in a layered superconductor in a parallel
     magnetic field.}
\end{figure}

We consider a strongly layered impure superconductor in a {\em parallel
magnetic field} [see Fig.~1]. The superconducting layers provide a
sufficiently strong confining potential for the (Josephson-like) vortex
lines which exist in the interlayer spacing. We shall exclude the
possibility of the lines crossing the superconducting layers. [For fields
parallel to the ab-planes of the Bi-compound, typical vortex kink energies
are of the order $10^3(1-T/T_c)^{^{\circ }}K$.] This amounts to limiting the
vortex displacement field from two components in an isotropic sample to one
component (i.e., parallel to the layers). For simplicity we focus our study
mostly on the {\em dilute} limit where the inter-vortex spacings $l_{\perp }$
(inter-layer), $l_{\Vert }$ (intra-layer) exceed the magnetic penetration
depths $\lambda _{ab}$, $\lambda _c$ respectively. The implications of our
results on the dense limit are straightforward and will be discussed below.

A well-established analytic description for a {\em single} layer of vortex
lines (for length scales exceeding $l_{\Vert }$) is given by the Hamiltonian
\cite{F}
\begin{equation}
\beta {\cal H}_{{\rm 2D}}[\phi _j,W_j]=\int_{{\bf r}_{\Vert }}\left\{ \frac K%
2(\nabla _{\Vert }\phi _j)^2-W_j[\phi _j({\bf r}_{\Vert }),{\bf r}_{\Vert
}]\right\} ,  \label{H2D}
\end{equation}
where $\phi _j({\bf r}_{\Vert })$ describes the in-plane displacement of the
vortex lines in the $j^{{\rm th}}$ layer and $K$ is an (isotropized)
in-plane elastic constant. Pinning effects due to point disorder are
described by the random potential $W_j[\phi _j({\bf r}_{\Vert }),{\bf r}%
_{\Vert }]$, with the second moment
\begin{equation}
\overline{W_j[\phi _j,{\bf r}_{\Vert }]W_j[\phi _j^{\prime },{\bf r}_{\Vert
}^{\prime }]}=g_0^2\cos [\phi -\phi ^{\prime }]\delta ^2({\bf r}_{\Vert }-%
{\bf r}_{\Vert }^{\prime }),  \label{disorder}
\end{equation}
where the overbar denotes disorder average, $g_0$ characterizes the (bare)
strength of the random potentials, and the cosine captures the {\em discrete}
nature of the vortex lines~\cite{natt1}. With many layers stacked next to each
other, the Hamiltonian for the whole system is
\begin{equation}
\beta {\cal H}=\sum_j\left\{ \beta {\cal H}_{{\rm 2D}}+\int_{{\bf r}_{\Vert
}}V_j[\phi _{j+1}({\bf r}_{\Vert })-\phi _j({\bf r}_{\Vert })]\right\} ,
\label{H3D}
\end{equation}
where $\overline{W_jW_{j^{\prime }}}=\delta _{j,j^{\prime }}\overline{W_jW_j}
$ since the bare random potentials in different layers are uncorrelated, and
the interaction has the form
\begin{equation}
V_j[\phi ]= - \mu \cos [\phi ].  \label{V}
\end{equation}
The expression (\ref{V}) can be regarded as the repulsive magnetic
interaction energy between the lowest harmonics of density fluctuations
between vortex lines in ``adjacent'' layers, a valid approximation in the
dilute limit \cite{MK}. The coupling constant $\mu $ is related to the shear
modulus of the flux line lattice. The main feature of (\ref{V}) is that it
goes beyond the elastic approximation, as it allows for dislocation loops
between adjacent layers [see Fig.~2].

\begin{figure}
\epsfxsize=3.2truein
\hskip 0.0truein \epsffile{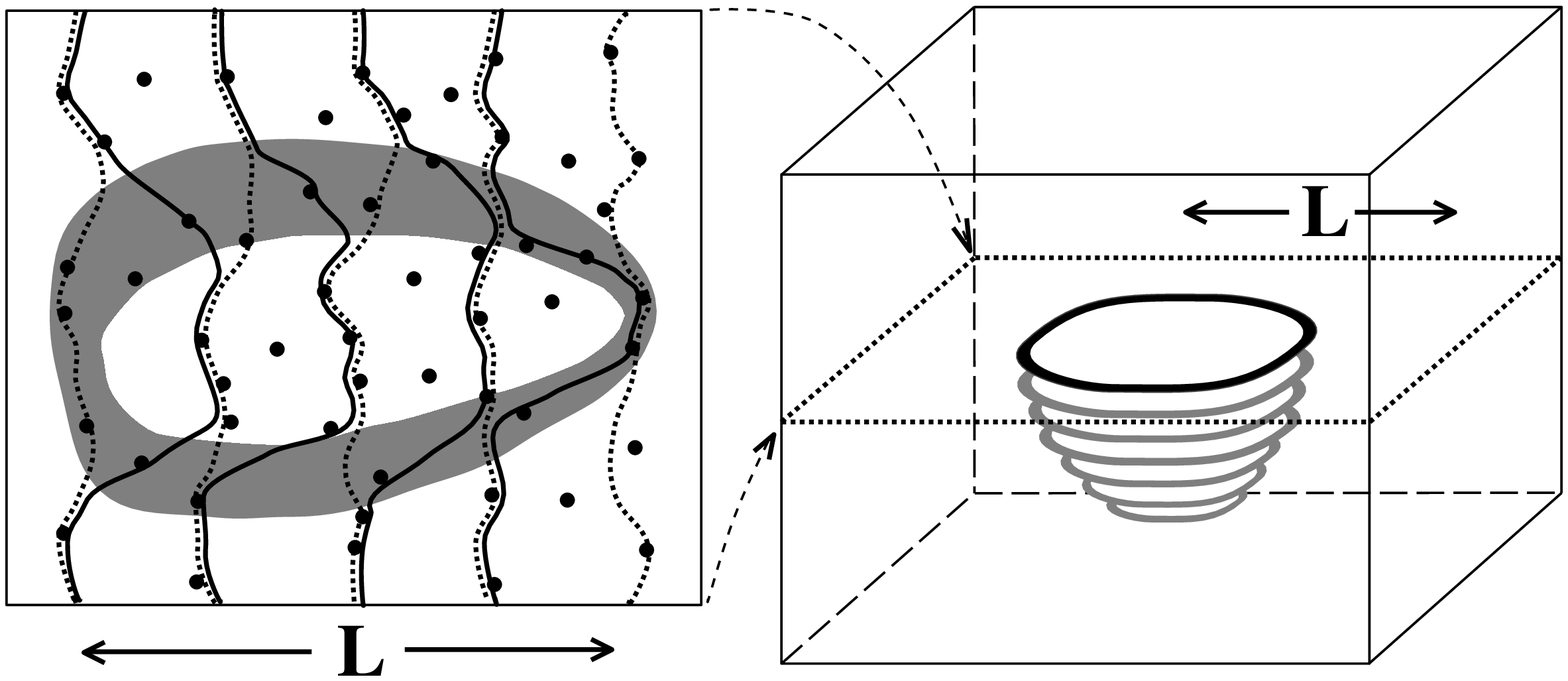}
\vspace{\baselineskip}
 { FIG 2: Elastic rearrangement of each planar flux array can be
represented by a number of vortex {\em loops}. An example is
shown on the left (shaded  region). Aligned vortex loops in successive
layers form a vortex {\em sheet} as shown on the right.
The  boundary of the sheet (dark loop on top) is a
{\em dislocation loop}.}
\end{figure}

In what follows, we first study the phase diagram of the system in terms of
the parameters $\mu$ and $K$. If the vortex layers are uncoupled, i.e., $%
\mu=0$, then each layer undergoes separately a glass transition at a
critical value $K_c = 1/4\pi$ \cite{F,CO}, with a vanishing linear
resistivity in the low temperature phase ($K>K_c$) and ohmic behavior in the
high temperature phase ($K<K_c$) \cite{GS,TS}. The physical range of
interest corresponds to $K \gg K_c$. When coupling the layers together via
Eq.~(\ref{V}), a competition arises between the tendency of vortex lines in
each layer to lower their free energy by adapting to the disorder in their
own layer, against the tendency to minimize the interlayer coupling by
bringing neighboring vortex layers into registry. The limit of very weak
coupling ($\mu \ll 1$) may be studied using perturbation theory; it is
straightforward to find that weak coupling is irrelevant at large scales.

In the limit of very large coupling $\mu \to \infty $, the interaction
potential $V[\phi ]$ may be replaced by the quadratic form, $\frac \mu 2%
\phi^2$, which describes an elastic (i.e., dislocation-free)
coupling in the direction perpendicular to the layers. This is just the
anisotropic (and one-component) version of the VG considered previously in
Refs.~\cite{natt1,GL,VF}. After introducing a continuum description in terms
of ${\bf r}=({\bf r}_{\Vert },r_{\perp })$ and rescaling $r_{\perp }=(j\cdot
l_{\perp })\sqrt{\mu l_\perp^2 /K}$, we get an isotropic 3D {\em elastic}
Hamiltonian
\begin{equation}
\beta {\cal H}_{{\rm 3D}}=\int d^3{\bf r}\left\{ \frac \gamma 2(\nabla \phi
)^2-W[\phi ({\bf r}),{\bf r}]\right\}   \label{H3Del}
\end{equation}
with an effective elastic constant $\gamma =\sqrt{\mu K}$ and a random
potential $W[\phi ({\bf r}),{\bf r}]$ with $\overline{W[\phi ,{\bf r}]W[0,0]}%
=g^2\cos [\phi ]\delta ^3({\bf r})$, where $g^2=g_0^2\sqrt{\mu /K}$. From
various methods including position-space RG \cite{VF}, functional RG \cite
{GL}, and the variational Ansatz with replica-symmetry breaking \cite{GL,K},
one finds that the system (\ref{H3Del}) forms a glass phase with
\begin{equation}
\overline{\langle [\phi ({\bf r})-\phi ({\bf r}')]^2\rangle }_{{\rm 3D}}
=2A\log \left(|{\bf r}-{\bf r}^{\prime }|/\ell\right)  \label{corr3D}
\end{equation}
{\em beyond} the positional correlation length $\ell \sim \gamma ^2/g^2$,
with $A$ being a universal number of $O(1)$. [In this model, $\ell$ is
equivalent to the Larkin length, although in more realistic models
including higher harmonics of density fluctuations, $\ell$ exceeds the
Larkin length~\cite{GL4}.]The logarithmic fluctuation in displacement
leads to quasi-long-range order and  an (algebraic) Bragg peak at reciprocal
lattice vector $2\pi/l_{\Vert }$. This phase has since been referred to as the
``Bragg glass''~\cite{GL}. In
3D, there is a large elastic energy cost, of the order $(\gamma /2)\int d^3%
{\bf r}\overline{\langle (\nabla \phi )^2\rangle }_{{\rm 3D}}\sim \gamma L$,
for logarithmic fluctuations in a volume of the order $L^3$. This energy is
compensated by the disorder energy gained from the anomalous
displacement of the flux array. Thus
\begin{equation}
\Delta E\sim \gamma L  \label{DE}
\end{equation}
gives the sample-to-sample free energy fluctuation of the Bragg
glass. Also, due to the existence of a statistical symmetry\cite
{symmetry,hf}, it is known that the disorder-averaged responses of the system
to various elastic deformations are identical to those of the pure elastic
system, i.e., Eq.~(\ref{H3Del}) with $W=0$. In particular,
there is a non-zero response to shear.

Given the above properties of the Bragg glass, which exists so far only in
the unphysical limit $\mu \to \infty $, our first task is to determine whether
it can persist at a finite $\mu $, i.e., whether the system is stable to the
{\em spontaneous} formation of dislocation loops on length scales much
larger than the correlation length. To investigate this possibility, we divide
the system into two halves (within which the layers are elastically
coupled), and allow dislocation loops to form in the contact plane, say
between the $j_0^{{\rm th}}$ and $(j_0+1)^{{\rm th}}$ layers. Analytically,
this is implemented by using the following interaction energy in
Eq.~(\ref{H3D}):
\begin{eqnarray}
V'_j[\phi] &=\frac \mu 2 \phi^2\quad &{\rm if}\quad j\neq j_0  \nonumber \\
  &= - \mu^{\prime }\cos [\phi]\quad  &{\rm if}\quad j=j_0,
\label{disloc}
\end{eqnarray}
where $\mu ^{\prime }\approx \mu $ approximates (\ref{V}). It is instructive
to consider arbitrary values of $\mu ^{\prime }$. Let us compare the
free energy difference between the completely coupled and completely decoupled
limit. If the two halves of the system
are decoupled, then each forms a Bragg glass, and the configuration of the
flux array in each half is {\em individually optimized}. But if the two halves
are tied together, then the constraint across the contact plane forces a
global
re-optimization of the flux array, resulting in a {\em higher} free energy
for each half. The typical free energy
increase due to the constraint is given by the sample-to-sample free energy
fluctuation of the Bragg glass, $\Delta E\sim \gamma L$. The expression (\ref
{DE}) is used here because each half of the system must be {\em completely}
re-optimized given the constraint, as if the configuration of the random
potential has been completely changed. The difference in the optimal
configuration of the flux array in each half resulting from the constraint
at $j_0$ can be described by a collection of ``vortex sheets'' such as the
one depicted in Fig.~2. The boundary of a vortex sheet is a dislocation
loop; it describes phase mismatches across the contact plane.

To find whether or not the two half systems actually couple for a given
$\mu'$, it is necessary to balance the cost in disorder energy due to
coupling with the {\em reduction}
 in interaction energy due to phase matching.
The latter can be readily computed for small $\mu'$ using (\ref{corr3D}).
One finds $\mu' \int_{L^2}d^2{\bf r}_{\Vert }
\overline{\langle \cos [\phi_{j_0+1}-\phi _{j_0}]\rangle }_{{\rm 3D}}
\sim \mu'\ell^A L^{2-A}\lesssim L$ since $A\ge 1$~\cite{GL}.
As the disorder energy cost to coupling, $\Delta E\sim \gamma L$,
exceeds the interaction energy to be gained at small $\mu ^{\prime }$
and large $L$, the two half
systems will remain decoupled. In the large (but finite) $\mu ^{\prime }$
limit of interest, the perturbative result is no longer valid. Since the
energy cost of phase mismatch is large there, we consider the stability of
a {\em single} optimally-configured dislocation loop of extent
$L\gg \ell $ at the contact plane of the two half systems that are otherwise
elastically coupled [Fig.~2].
The energy cost of the core of the dislocation loop due to
the inter-layer interaction is extensive. For a stretched circular loop of
linear size $L$, we expect $E_{{\rm core}}\sim \mu ^{\prime }\ell L$.
Here,  $\ell $ appears as the
``thickness'' of the loop because the flux array is elastically coupled
at smaller scales at low temperatures~\cite{GL3}.
More generally, if we allow the dislocation loop to take on fractal shapes,
say with the total loop length scaling as $L^D$ for $L\gg \ell$
($D\ge 1$ being the fractal dimension), then the
core energy becomes $E_{{\rm core}}\sim \mu' \ell^{2-D} L^D$.
On the other hand, the disorder energy {\em gained} from the formation of
a dislocation loop of size $L$
is just the energy gained from the formation of
a vortex sheet $E_{sheet}(L)$, resulting from the {\em elastic deformation}
of the half-system [see Fig.~2].
$E_{sheet}(L)$ clearly cannot exceed $\Delta E(L)$
which is the disorder energy gained from {\em complete} optimal elastic
rearrangement of the half-system at scale $L$. Assuming scaling of vortex
sheet energies, $E_{sheet}(L)\sim {\gamma ^{\prime }}L^\omega $, it
follows that $\omega \le 1$.

The actual value of the exponent $\omega$ depends on the structure of
the dislocation loop we allow, i.e.,  on the fractal dimension $D$.
We expect that the upper bound ($\omega=1$) may only be reached if the
structure
of the associated vortex sheet is similar to those occurring {\em naturally}
in complete elastic rearrangement of the half system. The latter can be
deduced as follows:
Denote the difference in the configuration before/after rearrangement by
$\varphi({\bf r})$. The vortex sheets are then the equal-$\varphi$ contours
of $\varphi({\bf r})$,
and the associated dislocation loops are the contours of
$\varphi({\bf r}_\|,r_\perp=j_0 l_\perp)$. The relationship between a rough
``landscape'' and the fractal geometry of its contours have recently been
examined~\cite{kh}. For a  logarithmically-rough landscape $\varphi$
[see Eq.~(\ref{corr3D})],
an exact calculation yields $D=3/2$~\cite{kh,sd,note0}.
Thus, we expect $\omega\le1$ for $D=3/2$, and $\omega < 1$ for $D<3/2$.
The total energy of the dislocation loop
\begin{equation}
E_{\rm loop} = E_{\rm core}(L) - E_{\rm sheet} (L)
\sim \mu' \ell^{2-D} L^D - \gamma' L^{\omega(D)} \label{eloop}
\end{equation}
does not admit a stable solution with $L\gg \ell$ for large
$\mu' \approx \mu$.
Hence the Bragg glass is stable to the spontaneous
formation and proliferation of large dislocation loops.
This conclusion is further supported by a systematic renormalization-group
analysis, the details of which will be given elsewhere~\cite{long}.
The possibility of a {\em marginally} stable Bragg glass for weakly-disorder
sample was first suggested in Ref.~\cite{GL}, based on the  assertions
that $\Delta E \sim g L$
 and $E_{\rm core} \sim c L$, where $g$ is the bare disorder strength which
can be made arbitrary small and $c$ is a given number~\cite{GL3}.
The above analysis indicates that the dislocation loops
are much more {\em strongly suppressed}
at low temperatures  by the anomalously large core energy.

Next, we investigate the extent of the stability regime for the Bragg glass.
As already shown, the Bragg glass cannot exist in the limit of very weak
inter-layer coupling. We expect the maximum possible extent of the
stability regime to be the point where the disorder is so strong that
the correlation length $\ell$ becomes of the order of the average vortex
spacing $a$.
Beyond this point, a topologically-ordered flux array cannot even form at
the smallest scale, and the entire collective pinning picture breaks down.
To find the actual size of the stability regime for the layered system
(\ref{H3D}), we consider quasi two-dimensional in-plane fluctuations
i.e., on the {\em shortest} scale in the direction perpendicular to the
layers. Analytically, we apply a variational
treatment, with the variational Hamiltonian obtained by replacing the
interaction potential $V_j[\phi ]$ in (\ref{H3D}) by the quadratic form $%
\widetilde{V}_j[\phi ]=\frac{\widetilde{\mu }}2 \phi^2$.
$\widetilde{V}$ describes an elastic (i.e., dislocation-free) coupling
in the direction perpendicular to the layers.
The parameter $\widetilde{\mu }$ has the meaning of an effective shear
modulus and may be determined self-consistently within the variational
treatment. The minimization of the variational free energy with respect to $%
\widetilde{\mu }$ yields the self-consistency equation
\begin{eqnarray}
\widetilde{\mu }=\mu \overline{\langle \cos [\phi _{j+1}({\bf r}_{\Vert
})-\phi _j({\bf r}_{\Vert })]\rangle }_{\widetilde{{\cal H}}}.  \label{scon}
\end{eqnarray}
This is evaluated using a Gaussian approximation which can
be justified~\cite{long} in a controlled fashion and should be
reasonably accurate.
$\overline{\langle \phi \phi \rangle }_{\widetilde{{\cal H}}}$ contains
contributions from (i) the quasi 2D VG regime $\overline{\langle \phi \phi
\rangle }_{2D}$ which dominates for $\widetilde{\mu }\approx 0$; (ii) the 3D
VG regime $\overline{\langle \phi \phi \rangle }_{3D}$; and (iii) thermal
fluctuations on scales smaller than the correlation length $\ell$ for large
$\widetilde{\mu }$. Using Eq.~(\ref{corr3D})
for $\overline{\langle \phi \phi \rangle }_{3D}$,
and using $\overline{\langle \phi \phi \rangle }_{2D}=2(1+\alpha)\log (L)$
(where $\alpha\approx (l_{\Vert }g_0/K)^4$ from Refs.~\cite{GL,K}),
the following
results are obtained~\cite{note1}: The self-consistency equation has a stable
solution with nonzero shear modulus only for~\cite{note2}
$\mu >\mu _c=c^2 K/\ell ^2$.
For $\mu < \mu_c$, the system ``melts'' into a stack of decoupled 2D VG's,
distinguished from the Bragg glass by a vanishing shear modulus
$\widetilde{\mu}=0$~\cite{melt}. The transition at $\mu =\mu _c$ is
first order, with a jump in $\widetilde{\mu }$~\cite{long}.
Our variational calculation yields a prefactor
$c \approx 5$ which depends very weakly on temperatures, as long as we are
away from the melting temperature of the pure system~\cite{long}.

It is illustrative to express $\mu$ and $K$ in terms of
the correlation length [of the anisotropic system (\ref{H3D})]
in the $\perp$-direction, $\ell_\perp \equiv \sqrt{\mu l_\perp^2/K} \ell$.
The above stability condition becomes
\begin{equation}
\ell_\perp >  c \cdot l_\perp. \label{stab_cond}
\end{equation}
Thus, we find the Bragg glass to be very stable, reaching within a factor
of 10 of the maximum extent of stability (i.e., $\ell \approx l$)
beyond which the very concept of collective pinning breaks down!
As variational calculations tend to under-estimate fluctuations, we
expect the actual value of $c$ to be somewhat larger.
Nevertheless, the condition (\ref{stab_cond}) appears to be quite general,
and may be viewed as the disordered-analog of the Lindemann criterion.

Applying this Lindemann-like criterion, we can expect the following
low-temperature  behaviors for the usual experimental situation of flux lines
perpendicular to the CuO planes; a detailed account
including the relevance to recent experiments will be
given elsewhere~\cite{long}. (a) In the dilute limit,
the Bragg glass phase is stable if the transverse correlation length
$\ell(B)$ exceeds (a few times) the inter-vortex spacing $l(B)$. (b) The same
conclusion is obtained in the dense limit [$\lambda \gg l$]
if $\ell \gtrsim \lambda$. (c) In the opposite limit of $l \lesssim
\ell \ll \lambda \to \infty$, the long-ranged magnetic
interaction leads to a much stronger glass phase,
with $\Delta E \sim L^2$~\cite{GL}, if the dislocation loops are forbidden.
This however provides a much larger disorder energy
to be gained from the formation of dislocation loops [i.e.,Eq.~(\ref{eloop})
with $\omega=2$], leading to the proliferation of large loops.
This regime may belong to the universality class of
the (long-ranged) Gauge glass~\cite{gauge}, or it may simply be a viscous
line liquid. (d) For the physical case of finite $\lambda$'s,
dislocation loops can form at scales $\lambda$ and below, and we
expect the system to remain topologically-ordered at large scales.

In summary, we find strong evidence supporting the existence of a
topologically-ordered vortex glass for the layered superconductors.
The stability regime is surprisingly large and is
given by a Lindemann-like criterion. Beyond this regime, the system
is dominated by {\em strong disorders}, and its properties are not known.
It may simply {\em melt} into a viscous line liquid,
or it may form another type of glass phase~\cite{GL}.

We are grateful to many helpful discussions with D.~S.~Fisher, T.~Giamarchi,
J.~Kondev, P.~Le~Doussal, D.~R.~Nelson, L.~H.~Tang, and I.~Lyuksyutov.
TH acknowledges the
support of A.P. Sloan Foundation and an ONR Young Investigator Award through
ONR-N00014-95-1-1002. TN acknowledges the hospitality of Harvard University
where part of the work was done, as well as the financial support of the
Volkswagen Foundation.

%
%

\end{multicols}

\begin{references}
\bibitem{note1}  Qualitatively similar results are obtained using
$\overline{\langle \phi \phi \rangle }_{2D}\sim \log ^2(L)$ as
found in \cite{CO}, provided that we are in the low temperature regime
$K \gg K_c$.
\bibitem{note2}  The microscopic length scale which appears in the
calculation is actually the 2D correlation length $\ell_{\rm 2D}=K/g_0$.
However at $\mu=\mu _c$, $\ell_{\rm 2D}\sim \ell/\sqrt{c}$~\cite{long}.
\bibitem{melt} The behaviors for $\mu \le \mu_c$ are probably peculiar to
the specific model (\ref{H3D}) considered and are not the focus
of this study.
\end{references}
\end{document}